\documentclass[12pt]{article}
\usepackage{amsmath}
\usepackage{graphicx,psfrag,epsf}
\usepackage{enumerate}
\usepackage{natbib}
\usepackage{url} 
\usepackage{amsthm}

\usepackage{amsfonts}
\usepackage{multirow, booktabs}
\usepackage[table]{xcolor}
\usepackage{cleveref}
\usepackage[ruled,vlined]{algorithm2e}
\usepackage[T1]{fontenc}
\usepackage[utf8]{inputenc}
\usepackage{authblk}
\usepackage[multiple]{footmisc}
\usepackage{blindtext,titlefoot}
\usepackage{sectsty}

\usepackage{amsmath}
\usepackage{amssymb}
\usepackage{amsfonts}
\usepackage{multirow}
\usepackage{amsthm}
\usepackage{url}
\usepackage{xcolor}
\usepackage{float}

\floatname{algorithm}{Procedure}

\newtheorem{theorem}{Theorem}

\newtheorem{corollary}{Corollary}

\theoremstyle{definition}

\newtheorem{example}{Example}
\newtheorem{remark}{Remark}

\makeatletter
\newcommand*{\rom}[1]{\expandafter\@slowromancap\romannumeral #1@}
\makeatother

\begin{document}

\sectionfont{\bfseries\large\sffamily}%
%

\subsectionfont{\bfseries\sffamily\normalsize}%
%


\def\spacingset#1{\renewcommand{\baselinestretch}%
{#1}\small\normalsize} \spacingset{1}

\begin{center}
    \LARGE\bf Sharpening the Rosenbaum Sensitivity Bounds to Address Concerns About Interactions Between Observed and Unobserved Covariates
\end{center}

\begin{center}
   \large $\text{Siyu Heng}^{*} \text{ and Dylan S. Small}^{\dagger}$
\end{center}

\begin{center}
    \large \textit{University of Pennsylvania}
\end{center}

\let\thefootnote\relax\footnotetext{* Graduate Group in Applied Mathematics and Computational Science, University of Pennsylvania. E-mail: \textsf{siyuheng@sas.upenn.edu}}
\let\thefootnote\relax\footnotetext{$\dagger$ Department of Statistics, The Wharton School, University of Pennsylvania. E-mail: \textsf{dsmall@wharton.upenn.edu}}

\begin{abstract}
In observational studies, it is typically unrealistic to assume that treatments are randomly assigned, even conditional on adjusting for all observed covariates. Therefore, a sensitivity analysis is often needed to examine how hidden biases due to unobserved covariates would affect inferences on treatment effects. In matched observational studies where each treated unit is matched to one or multiple untreated controls for observed covariates, the Rosenbaum bounds sensitivity analysis is one of the most popular sensitivity analysis models. In this paper, we show that in the presence of interactions between observed and unobserved covariates, directly applying the Rosenbaum bounds will almost inevitably exaggerate the report of sensitivity of causal conclusions to hidden bias. We give sharper odds ratio bounds to fix this deficiency. We illustrate our new method through studying the effect of anger/hostility tendency on the risk of having heart problems.
\end{abstract}

\noindent%
{\it Keywords:}  Causal inference; Gene-environment interaction; Interaction terms; Matching; Observational studies; Sensitivity analysis.

\clearpage

\spacingset{1.45} 

\section{Introduction}

In a randomized experiment, units are randomly assigned to the treatment group or control groups, perhaps by flipping a coin. In contrast, in an observational study, treatments are not randomly assigned to units and differences between the outcomes of the treated and control groups can be a biased estimate of the true treatment effect because of baseline differences between the treated and control groups. Baseline differences that can be captured by observed covariates can often be removed by model-based adjustments or matching. Among these methods, matching has been extensively used as a nonparametric way of adjusting for the observed covariates in observational studies: each treated unit is matched to one or several controls (i.e., untreated units) on baseline observed covariates such that the treated units and controls are similar in measured confounders as they would be under a randomized experiment, and the comparisons are made within these matched sets \citep{rubin1973matching, rosenbaum2002observational, rosenbaum2010design, hansen2004full, stuart2010matching, zubizarreta2013effect, pimentel2015large, zubizarreta2017optimal}.

However, there is typically the concern that some important baseline covariates are unobserved, so that the treatment assignments may not be random within each matched set. A sensitivity analysis asks how a departure from random assignment of treatment would affect the causal conclusion drawn from a primary analysis that assumes the treatment is randomly assigned conditional on the observed covariates. Among various sensitivity analysis models in matched observational studies, the Rosenbaum bounds sensitivity analysis \citep{Rosenbaum1987, rosenbaum2002observational} is one of the most popular. The Rosenbaum bounds sensitivity analysis introduces a uniform sensitivity parameter $\Gamma\geq 1$ bounding the ratio of the odds of treatment within each matched set: the more $\Gamma$ departs from 1, the more the treatment assignment potentially departs from random assignment in each matched set. Then researchers typically look at the ``worst-case" p-value, which is defined as the largest p-value given the sensitivity parameter $\Gamma$ over all possible arrangements of unobserved covariates (i.e., unmeasured confounders) \citep{rosenbaum2002observational}. For examples of studies using the Rosenbaum bounds sensitivity analysis, see \citet{normand2001validating}, \citet{rosenbaum2002observational}, \citet{rosenbaum2004design}, \citet{heller2009split}, \citet{silber2009time}, \citet{stuart2013commentary}, \citet{zubizarreta2013effect}, \citet{hsu2015strong}, \citet{zubizarreta2016effect}, \citet{ertefaie2018quantitative}, \citet{fogarty2019studentized}, \citet{karmakar2019integrating}, \citet{zhao2019sensitivityvalue}, and \citet{zhang2020selecting}. Many other sensitivity analysis models also build on the Rosenbaum bounds sensitivity analysis (e.g., \citealp{gastwirth1998dual, ichino2008temporary, rosenbaum2009amplification, nattino2018model, fogarty2019extended}).

In this article, we show that in the presence of any interactions between the observed and unobserved covariates in the logit model of the treatment assignment probability, the Rosenbaum bounds will almost inevitably be loose for some of the matched sets. Interactions between observed and unobserved covariates commonly exist in observational studies. One such setting is the extensively-studied ``gene-environment interaction" (G$\times$E), where two different genotypes respond to environmental variation in different ways (\citealp{ottman1996gene, caspi2002role}). In many studies, such genotypes were not identified, measured, or publicly available, and should be considered as unobserved covariates that can interact with some observed environmental covariates \citep{perusse1999genotype}. Directly applying the Rosenbaum bounds sensitivity analysis in such settings can greatly exaggerate the sensitivity of the causal conclusion to hidden bias. To perform a more informative and less conservative sensitivity analysis in matched studies, we give sharper odds ratio bounds when there is concern about a possible interaction between an observed covariate and an unobserved covariate. We apply our new result to study the causal effect of anger/hostility tendency on the risk of having heart disease. The data used for this work is publicly available at \url{https://www.ssc.wisc.edu/wlsresearch/}. The code and codebook used for this work are available at \url{https://github.com/siyuheng/Sharpening-the-Rosenbaum-Bounds}.

\section{A brief review of the Rosenbaum bounds sensitivity analysis}\label{Sec: review of RBSA}

We briefly review the classical framework for the Rosenbaum bounds sensitivity analysis for a matched observational study in which each treated unit is matched to one or more controls \citep{rosenbaum2002observational}. There are $I$ matched sets $i=1,\dots, I$, and matched set $i$ contains $n_{i}$ ($n_{i} \geq 2$) units, so $N=\sum_{i=1}^{I}n_{i}$ units in total. In each matched set, one unit received treatment and the others received control. Let $Z_{ij}=1$ if unit $j$ in matched set $i$ received treatment, otherwise let $Z_{ij}=0$. Therefore, we have $\sum_{j=1}^{n_{i}}Z_{ij}=1$ for all $i$. Let $\mathbf{x}_{ij}=(x_{ij(1)},\dots, x_{ij(K)})^{T}$ denote the $K$ observed covariates and $u_{ij}$ an unobserved covariate of unit $j$ in matched set $i$. The sets are matched for the observed covariates but not for the unobserved covariate, therefore $\mathbf{x}_{ij}=\mathbf{x}_{ij^{\prime}}$ for all $i,j$ and $j^{\prime}$, but possibly $u_{ij}\neq u_{ij^{\prime}}$ if $j \neq j^{\prime}$ \citep{rosenbaum2002observational}. Denote the common observed covariates for units in matched sets $i$ as $\mathbf{x}_{i}=(x_{i(1)}, \dots, x_{i(K)})^{T}$, where $\mathbf{x}_{i}=\mathbf{x}_{ij}=\mathbf{x}_{ij^{\prime}}$ for all $i, j, j^{\prime}$. Under the potential outcomes framework, if unit $j$ in matched set $i$ received treatment (i.e., $Z_{ij}=1$), we observe the potential outcome $r_{Tij}$; otherwise (i.e., $Z_{ij}=0$), we observe the potential outcome $r_{Cij}$ \citep{neyman1923application,rubin1974estimating}. Therefore, the observed outcome for each $i, j$ is $R_{ij}=Z_{ij}r_{Tij}+(1-Z_{ij})r_{Cij}$. Denote the collection of observed outcomes as $\mathbf{R}=(R_{11}, \dots, R_{In_{I}})^{T}$ and the collection of unobserved covariates as $\mathbf{u}=(u_{11},\dots,u_{In_{I}})^{T}$. Write $\mathcal{F}=\{(r_{Tij}, r_{Cij}, \mathbf{x}_{ij}, u_{ij}), \ i=1,\dots,I, \ j=1,\dots,n_{i}\}$, and let $\mathcal{Z}$ be the set of all possible values of $\mathbf{Z}=(Z_{11},\dots, Z_{In_{I}})^{T}$ where $\mathbf{Z} \in \mathcal{Z}$ if and only if $\sum_{j=1}^{n_{i}}Z_{ij}=1$ for all $i$. Let $|A|$ denote the number of elements of a finite set $A$, and define the indicator function $\mathbf{1}\{A\}=1$ if $A$ is true, and $\mathbf{1}\{A\}=0$ otherwise. Let $a\gg b$ denote that $a$ is much greater than $b$.

Fisher's sharp null hypothesis of no treatment effect asserts that $H_{0}: r_{Tij}=r_{Cij}$, for all $i, j$. In a randomized experiment, we can assume that $\text{pr}(\mathbf{Z}=\mathbf{z}\mid \mathcal{F}, \mathcal{Z})=1/|\mathcal{Z}|=1/(\prod_{i=1}^{I}n_{i})$ for all $\mathbf{z} \in \mathcal{Z}$. In a stratified randomized experiment with one unit in each matched set being randomly assigned to treatment, the significance level of a test statistic $T$ being greater than or equal to the observed value $t$ can be computed as 
\begin{align*}
    \text{pr}(T \geq t\mid \mathcal{F}, \mathcal{Z})&=\sum_{\mathbf{z}\in \mathcal{Z}}\mathbf{1}\{T(\mathbf{z}, \mathbf{R}) \geq t\} \ \text{pr}(\mathbf{Z}=\mathbf{z}\mid \mathcal{F}, \mathcal{Z})\\
    &=\frac{|\{\mathbf{z}\in \mathcal{Z}: T(\mathbf{z}, \mathbf{R}) \geq t \}|}{|\mathcal{Z}|}.
\end{align*}

In an observational study, it is often unrealistic to assume that treatment was randomly assigned, even within a matched set of units with the same observed covariates, due to the possible presence of an unobserved covariate. A sensitivity analysis is therefore needed to determine how departures from random assignment of treatment would affect the causal conclusions drawn from a primary analysis that assumes the treatment is randomly assigned within each matched set. Let $\pi_{ij}=P(Z_{ij}=1\mid \mathcal{F})$ denote the probability that unit $j$ in matched set $i$ will receive treatment. The Rosenbaum bounds sensitivity analysis considers that two units $ij$ and $ij^{\prime}$ in the same matched set $i$, with the same observed covariates $\mathbf{x}_{ij}=\mathbf{x}_{ij^{\prime}}=\mathbf{x}_{i}$, may differ in their odds of receiving the treatment by at most a factor of $\Gamma  \geq 1$:
\begin{equation}\label{eqn: sens model gamma}
\Gamma^{-1} \leq \frac{\pi_{ij}(1-\pi_{ij^{\prime}})}{\pi_{ij^{\prime}}(1-\pi_{ij})}\leq \Gamma, \quad \text{for all $i\in \{1,\dots,I\}$ and $j, j^{\prime}\in \{1,\dots,n_{i}\}$}.
\end{equation}
Constraint (\ref{eqn: sens model gamma}) is also known as the Rosenbaum bounds \citep{diprete20047}. It is clear that the more $\Gamma$ departs from 1, the more the treatment assignment potentially departs from random assignment. In the Rosenbaum bounds sensitivity analysis, people are interested in the ``worst-case" (i.e., the largest possible) p-value reported by a test statistic $T$ given its observed value $t$ under constraint (\ref{eqn: sens model gamma}) \citep{rosenbaum2002observational}. In practice, researchers gradually increase the sensitivity parameter $\Gamma$, report the ``worst-case" p-value under each $\Gamma$, and find the largest $\Gamma$ such that the ``worst-case" p-value exceeds the prespecified level $\alpha$. Such a changepoint $\Gamma$ is called ``sensitivity value" and informs the magnitude of potential hidden bias required to alter the causal conclusion \citep{zhao2019sensitivityvalue}.

For example, in a paired study where $n_{i}=2$ for all $i$, a commonly used family of test statistics are sign-score statistics, including McNemar's test and Wilcoxon's signed rank test. Their general form is $T_{\text{ss}}=\sum_{i=1}^{I}d_{i}\sum_{j=1}^{2}c_{ij}Z_{ij}$, where both $d_{i}\geq 0$ and $c_{ij}\in \{0,1\}$ are functions of $\mathbf{R}$ and so are fixed under $H_{0}$. When each $R_{ij}$ is binary, setting $d_{i}=1$ and $c_{ij}=R_{ij}$ gives McNemar's test. For $i=1,\dots,I$, define $\overline{T}_{\Gamma, i}$ to be independent random variables taking the value $d_{i}$ with probability $p_{i}^{+}$ and the value zero with probability $1-p_{i}^{+}$, where 
\[ p_{i}^{+}=\left\{
\begin{array}{ll}
      0 & \quad \text{if\ $c_{i1}=c_{i2}=0$}, \\
      1 & \quad  \text{if\ $c_{i1}=c_{i2}=1$}, \\
      \frac{\Gamma}{1+\Gamma} & \quad \text{if \ $c_{i1}\neq c_{i2}$}. \\
\end{array}
\right. \]
As shown in Section 4.3 in \citet{rosenbaum2002observational}, under the Rosenbaum bounds (\ref{eqn: sens model gamma}), for all $t$ and $\Gamma\geq 1$, we have $\text{pr}(T_{\text{ss}} \geq t\mid \mathcal{F}, \mathcal{Z})\leq \text{pr}(\sum_{i=1}^{I}\overline{T}_{\Gamma, i} \geq t \mid  \mathcal{F}, \mathcal{Z})$. That is, in the Rosenbaum bounds sensitivity analysis, the ``worst-case" p-value under $\Gamma$ reported by $T_{\text{ss}}$ given $t$ is $\text{pr}(\sum_{i=1}^{I}\overline{T}_{\Gamma, i} \geq t \mid  \mathcal{F}, \mathcal{Z})$. Assuming no interactions between observed and unobserved covariates, the upper bound $\text{pr}(\sum_{i=1}^{I}\overline{T}_{\Gamma, i} \geq t \mid  \mathcal{F}, \mathcal{Z})$ is sharp in the sense that it can be achieved for some arrangements of unobserved covariates; see Section 4.3 in \citet{rosenbaum2002observational}. However, as we will show in Theorem~\ref{Theorem: Rosenbaum bounds are loose} in Section~\ref{sec: The Rosenbaum bounds are loose}, in the presence of interactions between observed and unobserved covariates, the upper bound $\text{pr}(\sum_{i=1}^{I}\overline{T}_{\Gamma, i} \geq t \mid  \mathcal{F}, \mathcal{Z})$ induced from the Rosenbaum bounds (\ref{eqn: sens model gamma}) is in general loose in the sense that it cannot be achieved for any arrangements of unobserved covariates. The result in Theorem~\ref{Theorem: Rosenbaum bounds are loose} also holds for general matching regimes (including pair matching, matching with multiple controls, and full matching).

\section{The Rosenbaum bounds are loose in the presence of X-U interactions}\label{sec: The Rosenbaum bounds are loose}

The Rosenbaum bounds (\ref{eqn: sens model gamma}) is an odds ratio bound imposed on all matched sets that does not explicitly involve the observed covariates $\mathbf{x}_{ij}$ and a hypothesized unobserved covariate $u_{ij}$; it is natural to consider how it can be derived from a model expressed in terms of $\mathbf{x}_{ij}$ and $u_{ij}$ for the treatment assignment probability $\pi_{ij}$ \citep{rosenbaum2002observational}. Since the first paper on the Rosenbaum bounds sensitivity analysis \citep{Rosenbaum1987}, considering a logit form linking $\pi_{ij}$ to $\mathbf{x}_{ij}$ and $u_{ij}$ with no interactions between $\mathbf{x}_{ij}$ and $u_{ij}$ has been a routine way of interpreting the Rosenbaum bounds (\ref{eqn: sens model gamma}) and has been applied in numerous studies \citep{rosenbaum2002observational, diprete20047}:
\begin{equation}\label{eqn: logit model no interations}
    \log \big(\frac{\pi_{ij}}{1-\pi_{ij}}\big)=g(\mathbf{x}_{ij})+ \gamma u_{ij}, \quad u_{ij}\in [0,1],
\end{equation}
where $\gamma \in \mathbb{R}$ is unknown, and $g(\cdot)$ is an arbitrary unknown function of $\mathbf{x}_{ij}$. Note that the constraint $u_{ij}\in [0,1]$ is no more restrictive than assuming a bounded support of $u_{ij}$ and is only imposed to make the scalar $\gamma$ more interpretable \citep{Rosenbaum1987,rosenbaum2002observational}. For example, if the original unobserved covariate $\widetilde{u}_{ij}\in [0, \xi]$ for some $0<\xi<+\infty$, we just need to consider a normalized unobserved covariate $u_{ij}=\xi^{-1}\widetilde{u}_{ij}\in [0,1]$ and consider the logit model of $\pi_{ij}$ in terms of $\mathbf{x}_{ij}$ and $u_{ij}=\xi^{-1}\widetilde{u}_{ij}\in [0,1]$. It is then straightforward to show that the Rosenbaum bounds (\ref{eqn: sens model gamma}) can be implied from (\ref{eqn: logit model no interations}) with $\Gamma=\exp(|\gamma|)$ \citep{rosenbaum2002observational, diprete20047}. Note that the Rosenbaum bounds (\ref{eqn: sens model gamma}) are a type of nonparametric odds ratio bounds which can be implied from more general models on $\pi_{ij}$ than model (\ref{eqn: logit model no interations}) with appropriate equation linking $\Gamma$ and the proposed model on $\pi_{ij}$. In the previous literature, model (\ref{eqn: logit model no interations}) is extensively used as a working model for interpreting the Rosenbaum bounds due to its simplicity and clarity. However, as we will show in the rest of this section, only considering model (\ref{eqn: logit model no interations}) and ignoring potential interactions between observed and unobserved covariates can be harmful in a sensitivity analysis.

In this section, we instead consider a more general model of $\pi_{ij}$ in terms of $\mathbf{x}_{ij}$ and $u_{ij}$ allowing for any possible additive two-way interactions between each $\mathbf{x}_{ij(k)}$ and $u_{ij}$ ($X$-$U$ interactions):
\begin{equation}\label{eqn: general interraction logit model}
    \log \big(\frac{\pi_{ij}}{1-\pi_{ij}}\big)=g(\mathbf{x}_{ij})+\mathbf{\beta}^{T}\mathbf{x}_{ij}\times u_{ij}+ \gamma u_{ij}, \quad u_{ij}\in [0,1],
\end{equation}
where $\mathbf{\beta}^{T}\in \mathbb{R}^{K}$ and $\gamma \in \mathbb{R}$ are unknown, and $g(\cdot)$ is an arbitrary unknown function of $\mathbf{x}_{ij}$. Similar to the arguments under model (\ref{eqn: logit model no interations}), the constraint $u_{ij}\in [0,1]$ is no more restrictive than assuming a bounded support of $u_{ij}$ and is only imposed to make $\mathbf{\beta}^{T}$ and $\gamma$ more interpretable. When $\mathbf{\beta}^{T}=\mathbf{0}$ (i.e., no $X$-$U$ interactions), model (\ref{eqn: general interraction logit model}) reduces to the original model (\ref{eqn: logit model no interations}) that motivated the Rosenbaum bounds sensitivity analysis \citep{Rosenbaum1987}. Under (\ref{eqn: general interraction logit model}), according to the definition of $\Gamma$ in the Rosenbaum bounds (\ref{eqn: sens model gamma}), the following equation linking the prespecified sensitivity parameter $\Gamma$ and the unknown parameters $(\mathbf{\beta}^{T}, \gamma)$ can be obtained:
\begin{align}\label{eqn: expression of gamma}
    \Gamma&=\max_{i, j, j^{\prime}}\frac{\pi_{ij}(1-\pi_{ij^{\prime}})}{\pi_{ij^{\prime}}(1-\pi_{ij})}\quad \text{subject to $\mathbf{x}_{ij}=\mathbf{x}_{ij^{\prime}}$ and $u_{ij}, u_{ij^{\prime}} \in [0,1]$ for all $i, j, j^{\prime}$}\nonumber \\ &=\max_{i=1,\dots,I}\exp(|\mathbf{\beta}^{T}\mathbf{x}_{i}+\gamma|).
\end{align}
See Appendix for a derivation of equation (\ref{eqn: expression of gamma}). Note that when $\mathbf{\beta}^{T}=\mathbf{0}$, equation (\ref{eqn: expression of gamma}) reduces to the commonly used equation $\Gamma=\exp(|\gamma|)$ obtained under model (\ref{eqn: logit model no interations}). A key insight from equation (\ref{eqn: expression of gamma}) is that, in the presence of $X$-$U$ interactions, setting the sensitivity parameter $\Gamma$ not only incorporates our prior belief on the unknown structural parameters $(\mathbf{\beta}^{T}, \gamma)$, but also information about the matched observed covariates $\mathbf{x}_{i}$, $i=1,\dots,I$. The following result claims that the Rosenbaum bounds (\ref{eqn: sens model gamma}) will almost inevitably be conservative if there are any interactions between observed and unobserved covariates. 

\begin{theorem}\label{Theorem: Rosenbaum bounds are loose}
Consider the sensitivity parameter $\Gamma$ defined in the Rosenbaum bounds (\ref{eqn: sens model gamma}). Let $\Gamma>1$, and suppose that there exist two matched sets $i_{1}$ and $i_{2}$ such that $\mathbf{x}_{i_{1}}\neq \pm \mathbf{x}_{i_{2}}$. Then we have under model (\ref{eqn: general interraction logit model}), there exist some $\mathbf{x}^{*}\in \mathbb{R}^{K}$ and a subset $E$ of $\mathbb{R}^{K}$ of Lebesgue measure zero, such that for any $\mathbf{\beta}^{T}\neq \mathbf{0}$ (i.e., if there exist any interaction terms between $\mathbf{x}_{ij}$ and $u_{ij}$) and $\mathbf{\beta}^{T} \notin E$, the Rosenbaum bounds (\ref{eqn: sens model gamma}) are loose for any matched set $i$ with $\mathbf{x}_{i}\neq \pm \mathbf{x}^{*}$, in the sense that for any matched set $i$ with $\mathbf{x}_{i}\neq \pm \mathbf{x}^{*}$ there exists some $\Upsilon_{i}< \Gamma$ such that 
\begin{equation*}
    \Upsilon_{i}^{-1}\leq \frac{\pi_{ij}(1-\pi_{ij^{\prime}})}{\pi_{ij^{\prime}}(1-\pi_{ij})}\leq \Upsilon_{i}, \quad \text{for all $j, j^{\prime}$.}
\end{equation*}
\end{theorem}

Proofs of all theorems and corollaries in this article are in Appendix. We consider a simple example to illustrate the principle of Theorem~\ref{Theorem: Rosenbaum bounds are loose}. 
\begin{example}
Suppose that there is only one observed covariate $x_{ij} \in \{0,1\}$, and also an unobserved covariate $u_{ij}\in [0,1]$. Under model (\ref{eqn: general interraction logit model}), we have $\log(\frac{\pi_{ij}}{1-\pi_{ij}})=g(x_{ij})+\beta x_{ij}u_{ij}+ \gamma u_{ij}$. According to (\ref{eqn: expression of gamma}), we have $\Gamma=\max \{ \exp(|\gamma|), \exp(|\beta+\gamma|)\}$. It is clear that if $\beta \neq 0$ or $-2\gamma$, we have $\exp(|\gamma|)\neq \exp(|\beta+\gamma|)$. If $\Gamma= \exp(|\gamma|)>\exp(|\beta+\gamma|)$, then the Rosenbaum bounds (\ref{eqn: sens model gamma}) are loose for any matched set $i$ with the common observed covariate $x_{i}=1$. That is, for all matched set $i$ with $x_{i}=1$, we have
\begin{equation*}
    \Gamma^{-1} < \exp(-|\beta+\gamma|) \leq \frac{\pi_{ij}(1-\pi_{ij^{\prime}})}{\pi_{ij^{\prime}}(1-\pi_{ij})}\leq \exp(|\beta+\gamma|)< \Gamma, \ \ \text{for all $j, j^{\prime}$ and $u_{ij}, u_{ij^{\prime}} \in [0,1]$.}
\end{equation*}
Similarly, if $\Gamma=\exp(|\beta+\gamma|)>\exp(|\gamma|)$, the Rosenbaum bounds (\ref{eqn: sens model gamma}) are loose for any matched set $i$ with $x_{i}=0$. Therefore, when $\beta\neq 0$, unless $\beta \in \{-2\gamma\}$ (a subset of $\mathbb{R}$ of Lebesgue measure zero), the Rosenbaum bounds are loose for either all matched sets $i$ with $x_{i}=0$ or all matched sets $i$ with $x_{i}=1$. 
\end{example}

\section{Sharper odds ratio bounds accounting for X-U interactions}\label{sec: sharper bounds}

In this section, we give new odds ratio bounds that are sharper than the Rosenbaum bounds (\ref{eqn: sens model gamma}) when a researcher is concerned about the possible interaction between a particular observed covariate, say, the $k$th component $x_{(k)}$ of the observed covariates vector $\mathbf{x}$, and the unobserved covariate $u$. We consider a sub-model of (\ref{eqn: general interraction logit model}) which allows for possible interaction term linking $x_{(k)}$ and $u$:
\begin{equation}\label{eqn: logit model with single interaction}
    \text{logit}(\pi_{ij})=\log \big(\frac{\pi_{ij}}{1-\pi_{ij}}\big)=g(\mathbf{x}_{ij})+\widetilde{\beta}\ \widetilde{x}_{ij(k)} u_{ij}+ \gamma u_{ij}, \quad u_{ij}\in [0,1], 
\end{equation}
where $\widetilde{\beta}, \gamma \in \mathbb{R}$ are unknown, and $g(\cdot)$ is an unknown function of $\mathbf{x}_{ij}$. Again, the unobserved covariate $u_{ij}\in [0,1]$ is normalized to make $\widetilde{\beta}$ and $\gamma$ more interpretable. Each $\widetilde{x}_{ij(k)}=\frac{x_{ij(k)}-\min_{i, j}x_{ij(k)}}{\max_{i, j}x_{ij(k)}-\min_{i, j}x_{ij(k)}}\in [0,1]$ is also normalized to make $\widetilde{\beta}$ more interpretable. Again, note that when $\widetilde{\beta}=0$, model (\ref{eqn: logit model with single interaction}) reduces to the original model assuming no interaction terms that motivated the Rosenbaum bounds sensitivity analysis \citep{Rosenbaum1987}. In addition to the sensitivity parameter $\Gamma$ defined in (\ref{eqn: sens model gamma}) which quantifies the magnitude of the largest possible bias over all matched sets, when $\Gamma>1$, we introduce another prespecified sensitivity parameter $\lambda$ under model (\ref{eqn: logit model with single interaction}) as
\begin{equation}\label{def: lambda}
    \lambda=\frac{\frac{\partial \text{logit}(\pi_{ij})}{\partial u_{ij}}\mid x_{ij(k)}=\max_{i,j}x_{ij(k)}}{\frac{\partial \text{logit}(\pi_{ij})}{\partial u_{ij}}\mid  x_{ij(k)}=\min_{i,j}x_{ij(k)}}=\frac{\widetilde{\beta}+\gamma}{\gamma}, \quad \gamma\neq 0.
\end{equation}
The sensitivity parameter $\lambda$ quantifies how distinct the effects of $u$ on the treatment assignment probability can be under the largest and smallest possible values of $x_{(k)}$. Note that when $\widetilde{\beta}=0$ (i.e., no interaction between $x_{(k)}$ and $u$), we have $\lambda=1$. Let $\widetilde{x}_{i(k)}$ denote the normalized common covariate $x_{i(k)}$ for matched set $i$, therefore $\widetilde{x}_{i(k)}=\widetilde{x}_{ij(k)}=\widetilde{x}_{ij^{\prime}(k)}$ for all $j, j^{\prime}$. Then we have the following sharper odds ratio bounds.

\begin{theorem} \label{thm: improved sensi model gamma}
Consider the sensitivity parameter $\Gamma$ defined in the Rosenbaum bounds (\ref{eqn: sens model gamma}) with $\Gamma>1$. Under model (\ref{eqn: logit model with single interaction}) which allows for possible interaction between the observed covariate $x_{(k)}$ and the normalized unobserved covariate $u$, consider the sensitivity parameter $\lambda$ defined in (\ref{def: lambda}). Then we have
\begin{equation}\label{bounds: sharper}
	\Gamma_{\lambda, i}^{-1}\leq \frac{\pi_{ij}(1-\pi_{ij^{\prime}})}{\pi_{ij^{\prime}}(1-\pi_{ij})}\leq \Gamma_{\lambda, i} \quad \text{for all $i\in \{1,\dots,I\}$ and $j, j^{\prime}\in \{1,\dots,n_{i}\}$}, 
\end{equation}
where
\begin{equation*}
 \Gamma_{\lambda, i}=  \left\{
\begin{array}{ll}
      \Gamma^{|(\lambda-1) \widetilde{x}_{i(k)}+1|} &\quad \text{if\ $|\lambda| \leq 1$}, \\
      \Gamma^{|(1-\lambda^{-1}) \widetilde{x}_{i(k)}+\lambda^{-1}|} & \quad \text{if\ $|\lambda|>1$}. 
\end{array}
\right. 
\end{equation*}
We have $1\leq \Gamma_{\lambda, i}\leq \Gamma$ for all $i$, and the equality $\Gamma_{\lambda, i}=\Gamma$ holds for matched set $i$ if and only if at least one of the following three conditions holds: (a) $\lambda=1$; (b) $|\lambda|\leq 1$ and $x_{i(k)}=\min_{i}x_{i(k)}$; (c) $|\lambda|\geq 1$ and $x_{i(k)}=\max_{i}x_{i(k)}$. The bounds (\ref{bounds: sharper}) are sharp in the sense that for all $i, j, j^{\prime}$, there exist $u_{ij}, u_{ij^{\prime}} \in [0,1]$ such that $\{\pi_{ij}(1-\pi_{ij^{\prime}})\}/\{\pi_{ij^{\prime}}(1-\pi_{ij})\}=\Gamma_{\lambda, i}$.
\end{theorem}

A key feature of the sharper odds ratio bounds in Theorem~\ref{thm: improved sensi model gamma} is that they incorporate the information of the observed covariates among the matched samples, which is ignored by the Rosenbaum bounds (\ref{eqn: sens model gamma}). In Table~\ref{tab:sharper gamma}, we present some numerical illustrations of the gap between the sensitivity parameter $\Gamma$ in the Rosenbaum bounds (\ref{eqn: sens model gamma}) and the $\Gamma_{\lambda, i}$ in the sharper odds ratio bounds proposed in Theorem~\ref{thm: improved sensi model gamma}.

\begin{table}[H]
  \centering
  \caption{The $\Gamma_{\lambda, i}$ in the sharper odds ratio bounds proposed in Theorem~\ref{thm: improved sensi model gamma} under various $\Gamma$, $\lambda$, and $\widetilde{x}_{i(k)}\in [0,1]$. Note that $\Gamma_{\lambda, i}=\Gamma$ when $\lambda=1$. } 
  \label{tab:sharper gamma}
  \smallskip
  \smallskip
  \smallskip
  \small
  \centering
\begin{tabular}{cccccccc}
$\Gamma=2$ & $\lambda=\frac{1}{8}$ &  $\lambda=\frac{1}{4}$ &  $\lambda=\frac{1}{2}$ &  $\lambda=1$ &  $\lambda=2$ &  $\lambda=4$ &  $\lambda=8$  \\
$\widetilde{x}_{i(k)}=0$ & 2.00 & 2.00 & 2.00 & 2.00 & 1.41 & 1.19 & 1.09 \\
$\widetilde{x}_{i(k)}=\frac{1}{5}$ & 1.77 & 1.80 & 1.87 & 2.00 & 1.52 & 1.32 & 1.23 \\
$\widetilde{x}_{i(k)}=\frac{2}{5}$ & 1.57 & 1.62 & 1.74 & 2.00 & 1.62 & 1.46 & 1.39 \\
$\widetilde{x}_{i(k)}=\frac{3}{5}$ & 1.39 & 1.46 & 1.62 & 2.00 & 1.74 & 1.62 & 1.57 \\
$\widetilde{x}_{i(k)}=\frac{4}{5}$ & 1.23 & 1.32 & 1.52 & 2.00 & 1.87 & 1.80 & 1.77 \\
$\widetilde{x}_{i(k)}=1$ & 1.09 & 1.19 & 1.41 & 2.00 & 2.00 & 2.00 & 2.00 \\
$\Gamma=3$ & $\lambda=\frac{1}{8}$ &  $\lambda=\frac{1}{4}$ &  $\lambda=\frac{1}{2}$ &  $\lambda=1$ &  $\lambda=2$ &  $\lambda=4$ &  $\lambda=8$  \\
$\widetilde{x}_{i(k)}=0$ & 3.00 & 3.00 & 3.00 & 3.00 & 1.73 & 1.32 & 1.15 \\
$\widetilde{x}_{i(k)}=\frac{1}{5}$ & 2.48 & 2.54 & 2.69 & 3.00 & 1.93 & 1.55 & 1.39 \\
$\widetilde{x}_{i(k)}=\frac{2}{5}$ & 2.04 & 2.16 & 2.41 & 3.00 & 2.16 & 1.83 & 1.69 \\
$\widetilde{x}_{i(k)}=\frac{3}{5}$ & 1.69 & 1.83 & 2.16 & 3.00 & 2.41 & 2.16 & 2.04 \\
$\widetilde{x}_{i(k)}=\frac{4}{5}$ & 1.39 & 1.55 & 1.93 & 3.00 & 2.69 & 2.54 & 2.48 \\
$\widetilde{x}_{i(k)}=1$ & 1.15 & 1.32 & 1.73 & 3.00 & 3.00 & 3.00 & 3.00 
\end{tabular}
\end{table}

If the observed covariate $x_{(k)}\in \{0,1\}$ is a binary (dummy) variable, model (\ref{eqn: logit model with single interaction}) reduces to 
\begin{equation}\label{eqn: logit model with single interaction and binary covariate}
    \text{logit}(\pi_{ij})=g(\mathbf{x}_{ij})+\widetilde{\beta}\ x_{ij(k)} u_{ij}+ \gamma u_{ij}, \quad x_{ij(k)}\in \{0,1\}, u_{ij}\in [0,1], 
\end{equation}
and the sensitivity parameter $\lambda$ as defined in (\ref{def: lambda}) reduces to
\begin{equation*}
    \lambda=\frac{\frac{\partial \text{logit}(\pi_{ij})}{\partial u_{ij}}\mid x_{ij(k)}=1}{\frac{\partial \text{logit}(\pi_{ij})}{\partial u_{ij}}\mid  x_{ij(k)}=0}=\frac{\widetilde{\beta}+\gamma}{\gamma}, \quad \gamma \neq 0.
\end{equation*}
That is, the sensitivity parameter $\lambda$ is simply the ratio of the effect of $u$ on the logit of the treatment assignment probability (denoted as $\partial\text{logit}/\partial u$) conditional on $x_{(k)}=1$ to that conditional on $x_{(k)}=0$. Theorem~\ref{thm: improved sensi model gamma} implies the following sharper odds ratio bounds when $x_{(k)}$ is binary.

\begin{corollary}\label{corollary: dummy improved sensi model gamma}
    Under the same setting as that in Theorem~\ref{thm: improved sensi model gamma}, if the observed covariate $x_{(k)}\in \{0,1\}$ is a binary (dummy) variable, we have:
    \begin{enumerate}
            \item If $|\lambda|=1$, then the Rosenbaum bounds (\ref{eqn: sens model gamma}) are sharp for all matched sets, in the sense that for all $i,j,j^{\prime}$, there exist some $u_{ij}, u_{ij^{\prime}} \in [0,1]$ such that $\{\pi_{ij}(1-\pi_{ij^{\prime}})\}/\{\pi_{ij^{\prime}}(1-\pi_{ij})\}=\Gamma$ or $\Gamma^{-1}$.
    \item  If $|\lambda|<1 $, then the Rosenbaum bounds (\ref{eqn: sens model gamma}) are sharp for all matched sets $i$ with $x_{i(k)}=0$. While for all matched sets $i$ with $x_{i(k)}=1$, the Rosenbaum bounds (\ref{eqn: sens model gamma}) can be improved with: for all $i, j, j^{\prime}$ with $x_{i(k)}=1$, we have $\Gamma^{-|\lambda|}\leq \{\pi_{ij}(1-\pi_{ij^{\prime}})\}/\{\pi_{ij^{\prime}}(1-\pi_{ij})\} \leq \Gamma^{|\lambda|}$.
\item If $|\lambda|>1 $, then the Rosenbaum bounds (\ref{eqn: sens model gamma}) are sharp for all matched sets $i$ with $x_{i(k)}=1$. While for all matched sets $i$ with $x_{i(k)}=0$, the Rosenbaum bounds (\ref{eqn: sens model gamma}) can be improved with: for all $i, j, j^{\prime}$ with $x_{i(k)}=0$, we have $\Gamma^{-1/|\lambda|}\leq \{\pi_{ij}(1-\pi_{ij^{\prime}})\}/\{\pi_{ij^{\prime}}(1-\pi_{ij})\} \leq \Gamma^{1/|\lambda|}$.
\end{enumerate}
\end{corollary}

Corollary~\ref{corollary: dummy improved sensi model gamma} implies that in the binary covariate case, the sign of the sensitivity parameter $\lambda$ does not matter in a sensitivity analysis. It also implies that in this case the more $|\lambda|$ departs from $1$, the less the treatment assignments can potentially depart from random assignments within some matched sets. When $|\lambda|=1$, the bounds in Corollary~\ref{corollary: dummy improved sensi model gamma} reduce to the Rosenbaum bounds (\ref{eqn: sens model gamma}).

\begin{remark}\label{remark: approximation}
Note that even if some observed covariates are not exactly matched for some matched sets, the sharper odds ratio bounds proposed in Theorem~\ref{thm: improved sensi model gamma} and Corollary~\ref{corollary: dummy improved sensi model gamma} can still be used as approximate sensitivity bounds as long as the following three conditions hold: 1) the function $g(\mathbf{x}_{ij})$ in model (\ref{eqn: logit model with single interaction}) is a continuous function; 2) the observed covariates are very close among individuals within each matched set $i$ (i.e., $\mathbf{x}_{ij}\approx \mathbf{x}_{ij^{\prime}}$ for all $i, j, j^{\prime}$); 3) the observed covariates $x_{(k)}$ which can interact with the unobserved covariate are equal among individuals within each matched set (i.e., $x_{ij(k)}=x_{ij^{\prime}(k)}=x_{i(k)}$ for all $i, j, j^{\prime}$). Then the following calculation, combined with repeating the arguments in the proof of Theorem~\ref{thm: improved sensi model gamma} in Appendix, can justify the validity of the approximation: 
\begin{align*}
    \frac{\pi_{ij}(1-\pi_{ij^{\prime}})}{\pi_{ij^{\prime}}(1-\pi_{ij})}&=\frac{\exp\{ g(\mathbf{x}_{ij})+\widetilde{\beta}\ \widetilde{x}_{ij(k)} u_{ij}+ \gamma u_{ij}\}}{\exp\{ g(\mathbf{x}_{ij^{\prime}})+\widetilde{\beta}\ \widetilde{x}_{ij^{\prime}(k)} u_{ij^{\prime}}+ \gamma u_{ij^{\prime}}\}}\\
    &=\frac{\exp\{ g(\mathbf{x}_{ij})\}}{\exp\{ g(\mathbf{x}_{ij^{\prime}})\}}\frac{\exp\{ \widetilde{\beta}\ \widetilde{x}_{ij(k)} u_{ij}+ \gamma u_{ij}\}}{\exp\{ \widetilde{\beta}\ \widetilde{x}_{ij^{\prime}(k)} u_{ij^{\prime}}+ \gamma u_{ij^{\prime}}\}}\\
    &\approx \exp\{ (\widetilde{\beta}\ \widetilde{x}_{i(k)}+\gamma)(u_{ij}-u_{ij^{\prime}}) \} \quad (\text{since $\mathbf{x}_{ij}\approx \mathbf{x}_{ij^{\prime}}$ and $\widetilde{x}_{ij(k)}=\widetilde{x}_{ij^{\prime}(k)}=\widetilde{x}_{i(k)}$}).
\end{align*}

\end{remark}

\section{Implementation of the proposed sharper odds ratio bounds}

As we have shown in Theorem~\ref{thm: improved sensi model gamma} in Section~\ref{sec: sharper bounds}, taking potential $X$-$U$ interactions can lead to sharper odds ratio bounds than the Rosenbaum bounds (\ref{eqn: sens model gamma}). By implementing the sharper odds ratio bounds proposed in Theorem~\ref{thm: improved sensi model gamma}, researchers can perform a less conservative sensitivity analysis than that with directly applying the Rosenbaum bounds. The sharper odds ratio bounds proposed in Theorem~\ref{thm: improved sensi model gamma} can be directly embedded in many of the previous results in the Rosenbaum bounds sensitivity analysis: in many cases, researchers just need to follow the procedure of the Rosenbaum sensitivity analysis except replacing all the sensitivity parameter $\Gamma$ with $\Gamma_{\lambda, i}$ proposed in Theorem~\ref{thm: improved sensi model gamma} in each matched set. For example, the following result shows how applying Theorem~\ref{thm: improved sensi model gamma} to perform a sensitivity analysis with a sign-score statistic $T_{\text{ss}}$ can result in a less conservative ``worst-case" p-value than the one reported by directly applying the Rosenbaum bounds sensitivity analysis with $T_{\text{ss}}$.

\begin{corollary}\label{corollary: smaller worst-case p-value}
    Let $T_{\text{ss}}=\sum_{i=1}^{I}d_{i}\sum_{j=1}^{2}c_{ij}Z_{ij}$ be a sign-score statistic as introduced in Section~\ref{Sec: review of RBSA}. Consider testing Fisher's sharp null of no treatment effect $H_{0}$, and the sensitivity parameters $\Gamma$ defined in the Rosenbaum bounds (\ref{eqn: sens model gamma}) and $\lambda$ defined in (\ref{def: lambda}) under model (\ref{eqn: logit model with single interaction}). Define $\Gamma_{\lambda, i}$ as in Theorem~\ref{thm: improved sensi model gamma} and $\overline{T}_{\Gamma, i}$ as in Section~\ref{Sec: review of RBSA}. For $i=1,\dots,I$, define $\widetilde{T}_{\Gamma, \lambda, i}$ to be independent random variables taking the value $d_{i}$ with probability $\widetilde{p}_{\lambda, i}$ and the value zero with probability $1-\widetilde{p}_{\lambda, i}$, where 
\[ \widetilde{p}_{\lambda, i}=\left\{
\begin{array}{ll}
      0 & \quad \text{if\ $c_{i1}=c_{i2}=0$}, \\
      1 & \quad  \text{if\ $c_{i1}=c_{i2}=1$}, \\
      \frac{\Gamma_{\lambda, i}}{1+\Gamma_{\lambda, i}} & \quad \text{if \ $c_{i1}\neq c_{i2}$}. \\
\end{array}
\right. \]
    Then for all $t$ and any fixed $\Gamma> 1$ and $\lambda \in \mathbb{R}$, we have $\text{pr}(T_{\text{ss}} \geq t\mid \mathcal{F}, \mathcal{Z})\leq \text{pr}(\sum_{i=1}^{I}\widetilde{T}_{\Gamma, \lambda, i} \geq t \mid  \mathcal{F}, \mathcal{Z})$ for any $\mathbf{u}\in [0,1]^{N}$, and we have $ \text{pr}(\sum_{i=1}^{I}\widetilde{T}_{\Gamma, \lambda, i} \geq t \mid  \mathcal{F}, \mathcal{Z})\leq  \text{pr}(\sum_{i=1}^{I}\overline{T}_{\Gamma, i} \geq t \mid  \mathcal{F}, \mathcal{Z})$. The upper bound $ \text{pr}(\sum_{i=1}^{I}\widetilde{T}_{\Gamma, \lambda, i} \geq t \mid  \mathcal{F}, \mathcal{Z})$ is sharp in the sense that it can be achieved for some $\mathbf{u}$. 
\end{corollary}

\section{Illustration: the effect of anger/hostility tendency on heart problems}\label{Sec: Real data}

Type A behavior is characterized by hostility, intense ambition, competitive ``drive,” constant preoccupation with deadlines, and a sense of time urgency \citep{rosenman1976multivariate}. Early research data suggested type A behavior was related to heart problems but the original findings have not been supported by subsequent research \citep{myrtek2001meta}. Some researchers have turned their focus to whether tending to be angry and hostile -- one of the specific aspects of type A personality -- could cause heart problems \citep{chida2009association}. To study this, we consider data among males from the Wisconsin Longitudinal Study, a long-term study of a random sample of individuals graduated from Wisconsin high schools in 1957 \citep{herd2014cohort}. We define a binary indicator of tending to be angry/hostile (i.e., treated) if the respondent said on the 1992-1993 survey (when respondents were approximately 53) that in the last week there were three or more days on which the respondent felt angry or hostile for several hours and 0 (i.e., control) if there were no such day in the last week. We compare high anger/hostility tendency to low anger/hostility tendency and exclude middle levels of anger/hostility tendency because making the treated and control groups sharply differ in dose increases the insensitivity of a study to hidden bias \citep{rosenbaum2004design}. We take the outcome (heart problem indicator) to be 1 if the respondent reported having had a heart attack, coronary heart disease, or other heart problems in the 2003-2005 survey, and 0 otherwise. We pair match each treated individual with a control on the following cardiovascular disease risk factors \citep{kawachi1996prospective}: age, educational attainment, body mass indicator, drinking alcohol or not, smoking regularly or not, and childhood maltreatment indicator. The childhood maltreatment indicator is 1 if the respondent reported any childhood physical or sexual abuse, and 0 otherwise. Childhood maltreatment has been found to be associated with both anger/hostility tendency and heart problems \citep{carver2014adulthood, korkeila2010childhood}, and therefore is a confounder that needs to be controlled for. We discarded all the records with missing outcomes or covariates, and use optimal matching \citep{rosenbaum2010design} to match each treated with a control for the six baseline observed covariates, leaving 54 matched pairs. The absolute standardized differences (i.e., difference in means divided by the pooled standard deviation) between the treated and control groups are less than $0.1$ for all the six baseline observed covariates, indicating good overall balance \citep{rosenbaum2010design}. The smoking indicator and childhood maltreatment indicator are exactly matched between the treated unit and control within each matched pair. Although the other four covariates, i.e., age, educational attainment, body mass indicator, and drinking alcohol or not, are not exactly matched between the treated unit and its matched control for all pairs, they are closely matched for most pairs as the correlations for these four covariates equal 0.92, 0.98, 0.95, and 0.86 respectively. Therefore, according to Remark~\ref{remark: approximation}, in our study the sharper odds ratio bounds developed in Section~\ref{sec: sharper bounds} can be applied as approximate sensitivity bounds in our sensitivity analysis.

Another covariate we are concerned about as a confounder is the genotype monoamine oxidase A (MAOA) which has been found to be associated with both aggressive behavior and heart disease \citep{mcdermott2009monoamine, kaludercic2011monoamine}. The genetic data of the Wisconsin Longitudinal Study is not publicly available, therefore here we treat MAOA genotype as an unobserved covariate. We denote the unobserved MAOA genotype indicator (i.e., the $u$ in model (\ref{eqn: logit model with single interaction})) to be 1 if the individual has low-activity MAOA genotype (MAOA-L), and 0 if high-activity MAOA genotype (MAOA-H). According to a controlled experiment done in \citet{mcdermott2009monoamine}, individuals with MAOA-L are more likely to show aggression, suggesting $\gamma>0$ in model (\ref{eqn: logit model with single interaction}). Childhood maltreatment has been shown to significantly interact with MAOA genotype to confer risk for aggressive behavior: maltreated children with MAOA-L are more likely to develop violent behavior or show hostility \citep{caspi2002role,byrd2014maoa}, suggesting that the coefficient of the interaction term $\widetilde{\beta}$ in model (\ref{eqn: logit model with single interaction}) is greater than 0 and that the sensitivity parameter $\lambda=(\widetilde{\beta}+\gamma)/\gamma>1$. While setting a precise range for $\lambda$ needs further empirical study, some related studies suggest that $\lambda \gg 1$. For example, according to Figure 2A in \citet{caspi2002role}, among severely maltreated (during childhood) males, the logit of probability of conducting disorder among these with MAOA-L is much greater than that among those with MAOA-H. In contrast, among non-maltreated males, these two logits are extremely close. Therefore, if we treat the conducting disorder indicator as a proxy for the anger/hostility tendency indicator, results from \citet{caspi2002role} suggest that $\partial\text{logit}/\partial u $ if maltreated is much greater than that if non-maltreated (i.e., $\widetilde{\beta}+\gamma\gg \gamma$), implying $\lambda \gg 1$.

\begin{table}[ht]
  \centering
  \caption{The ``worst-case" p-values reported by McNemar's test under various $\Gamma$ and $\lambda$. When $|\lambda|=1$, they are the same as those reported by the Rosenbaum bounds sensitivity analysis.} 
  \label{tab:p-values}
  \smallskip
  \smallskip
  \smallskip
  \small
  \centering
\begin{tabular}{cccccccc}
& $|\lambda|=\frac{1}{8}$ &  $|\lambda|=\frac{1}{4}$ &  $|\lambda|=\frac{1}{2}$ &  $|\lambda|=1$ &  $|\lambda|=2$ &  $|\lambda|=4$ &  $|\lambda|=8$  \\
$\Gamma=1.31$ & 0.037 & 0.039 & 0.042 & \textbf{0.050} & 0.033 & 0.027 & 0.024\\
$\Gamma=1.37$ & 0.043 & 0.045 & \textbf{0.050} & 0.060 & 0.038 & 0.030 & 0.026 \\
$\Gamma=1.42$ & 0.048 & \textbf{0.050} & 0.056 & 0.068 & 0.042 & 0.032 & 0.028 \\
$\Gamma=1.44$ & \textbf{0.050} & 0.052 & 0.058 & 0.072 & 0.043 & 0.033 & 0.028 \\
$\Gamma=1.52$ & 0.058 & 0.061 & 0.069 & 0.087 & \textbf{0.050} & 0.036 & 0.031 \\
$\Gamma=1.81$ & 0.090 & 0.098 & 0.114 & 0.150 & 0.076 & \textbf{0.050} & 0.040 \\
$\Gamma=2.11$ & 0.127 & 0.140 & 0.166 & 0.223 & 0.105 & 0.065 & \textbf{0.050} \\
Sensitivity value & 1.44 & 1.42 & 1.37 & 1.31 & 1.52 & 1.81 & 2.11 
\end{tabular}
\end{table}

We use Corollary~\ref{corollary: smaller worst-case p-value} to calculate the ``worst-case" p-values $\text{pr}(\sum_{i=1}^{I}\widetilde{T}_{\Gamma, \lambda, i} \geq t \mid  \mathcal{F}, \mathcal{Z})$ reported by McNemar's test under various $\Gamma$ and $\lambda$, where $\lambda$ quantifies the possible interaction between the childhood maltreatment indicator and MAOA genotype; see Table~\ref{tab:p-values}. We also report corresponding sensitivity values under various $\lambda$. Note that when $|\lambda|=1$, the ``worst-case" p-values are the same as those reported by the Rosenbaum bounds sensitivity analysis. As discussed above, we are particularly concerned about the cases with $\lambda >1$. From Table~\ref{tab:p-values}, we can see that through applying the sharper odds ratio bounds developed in Section~\ref{sec: sharper bounds}, the ``worst-case" p-values are much less conservative than those reported by directly applying the Rosenbaum bounds sensitivity analysis, especially when $\lambda$ is much greater than 1 (i.e., there is a significant $X$-$U$ interaction), making a sensitivity analysis significantly more insensitive to hidden bias caused by the potential unobserved covariate. Therefore, for this particular data set, directly applying the Rosenbaum bounds sensitivity analysis can only detect a significant treatment effect up to a moderate magnitude of hidden bias (i.e., $\Gamma=1.31$). In contrast, applying our sharper odds ratio bounds to perform a sensitivity analysis allows the researcher to detect a significant treatment effect when there is a significant $X$-$U$ interaction, say, $\lambda\geq 2$, up to a significantly larger magnitude of hidden bias, $\Gamma=1.52$. A bias of $\Gamma=1.5$ is nontrivial as it corresponds to an unobserved covariate that doubles the odds of treatment and increases the odds of a positive treated-minus-control difference in observed outcomes by a factor of 4 \citep{rosenbaum2009amplification}.

\section{Discussion}

We here provide some practical guidance for empirical researchers on when and how our new odds ratio bounds should be used when conducting a Rosenbaum-type sensitivity analysis in matched studies. On the one hand, if a researcher has some prior knowledge about in which direction or to what extent the effects of the concerned unobserved covariate on the treatment assignment probability should vary with different values of the related observed covariate (i.e., a plausible range of the sensitivity parameter $\lambda$ defined in (\ref{def: lambda})), we strongly recommend that, instead of just conducting the sensitivity analysis using the traditional Rosenbaum bounds (\ref{eqn: sens model gamma}) (i.e., setting $\lambda=1$), she or he can also report the results of sensitivity analysis under a plausible range of $\lambda$ to better incorporate the expert knowledge to make the sensitivity analysis more informative and less conservative, as shown in Section~\ref{Sec: Real data}. On the other hand, even if there is no current evidence about the existence of any $X$-$U$ interactions or credible information on the range of $\lambda$ for the concerned $X$-$U$ interaction term, an empirical researcher could still benefit from our new methods. Suppose a researcher conducted the Rosenbaum bounds sensitivity analysis and found that the ``worst-case" p-values $> \alpha=0.05$ even under $\Gamma$ close to 1, i.e., the sensitivity value is small. Instead of rushing to claim that the causal conclusion is inevitably sensitive to hidden bias, the researcher can diagnose the reasons for sensitivity by selecting some candidate $X$-$U$ interaction terms and checking the corresponding ``worst-case" p-values and sensitivity values under various $\lambda$ through our new odds ratio bound. If the sensitivity values are always small for a reasonably wide range of $\lambda$, then she or he can confirm that the causal conclusion should indeed be sensitive to hidden bias regardless of potential $X$-$U$ interactions. If instead the sensitivity value becomes substantially larger as $\lambda$ departs from 1, then this implies that the previous finding that the causal conclusion is sensitive to hidden bias could be due to ignoring the possible $X$-$U$ interactions, in which case the researcher can do more investigation on the possibility of the actual existence of such $X$-$U$ interactions to report the sensitivity analysis in a more comprehensive way. 

There are limitations to the new odds ratio bound introduced in this work. First, it is only applicable for two-way $X$-$U$ interactions. For example, if there is an additional three-way interaction term $\widetilde{x}_{ij(k)}\widetilde{x}_{ij(k^{\prime})}u_{ij}$ in the treatment assignment probability model (\ref{eqn: logit model with single interaction}) for some $k\neq k^{\prime}$, then the sensitivity parameter $\lambda$ defined in (\ref{def: lambda}) cannot fully capture how the effects of the unobserved covariate $u_{ij}$ on the treatment assignment probability $\pi_{ij}$ would vary with different values of the two observed covariates $\widetilde{x}_{ij(k)}$ and $\widetilde{x}_{ij(k^{\prime})}$. Second, when the observed covariate $\widetilde{x}_{ij(k)}$ in the concerned $X$-$U$ interaction term is not binary, our new odds ratio bound is not applicable if the interaction term in model (\ref{eqn: logit model with single interaction}) is instead $f(\widetilde{x}_{ij(k)})u_{ij}$ where $f(\widetilde{x}_{ij(k)})$ is nonlinear in $\widetilde{x}_{ij(k)}$. Third, our new odds ratio bound cannot directly handle multiple $X$-$U$ interaction terms, e.g., when there are two interaction terms $\widetilde{x}_{ij(k)}u_{ij}$ and $\widetilde{x}_{ij(k^{\prime})}u_{ij}$ in model (\ref{eqn: logit model with single interaction}). Although in principle this type of problems can be solved by introducing additional sensitivity parameters into the odds ratio bounds, doing so can make a sensitivity analysis complicated and hard to interpret. Fourth, in the presence of $X$-$U$ interactions, after computing the sensitivity value, it requires further study to clarify how large the sensitivity value needs to be to make a study insensitive to hidden bias. Assuming no $X$-$U$ interactions, \citet{hsu2013calibrating} proposed a strategy that calibrates the values of the sensitivity parameters in matched observational studies to the observed covariates to help empirical researchers set plausible ranges of values for the sensitivity parameters. It would be helpful to investigate how to extend \citet{hsu2013calibrating}'s approach to the settings allowing $X$-$U$ interactions. Fifth, in this work, we only consider testing Fisher's sharp null hypothesis and have not discussed testing other types of null hypotheses such as Neyman's weak null hypothesis. In pair-matched observational studies, a promising direction for extending the proposed method in this article to allow testing Neyman's weak null hypothesis is to investigate if the proposed sharper odds ratio bounds proposed in Theorem~\ref{thm: improved sensi model gamma} can be incorporated into the studentized sensitivity analysis framework for the sample average treatment effect developed in \citet{fogarty2019studentized}. Despite these limitations, this work shows how investigating the confounding mechanism more carefully in a matched observational study can make a sensitivity analysis more informative and comprehensive. It might be fruitful for future research to explore how other structural constraints besides the one we explored might be used.

\section*{Acknowledgements}

The authors would like to thank the participants in the causal inference reading group of the University of Pennsylvania for helpful comments.
\par

\section*{Appendix}

\subsection*{A derivation of equation (\ref{eqn: expression of gamma})}

According to the definition of $\Gamma$ in the Rosenbaum bounds (\ref{eqn: sens model gamma}),
\begin{equation*}
    \Gamma=\max_{i, j, j^{\prime}}\frac{\pi_{ij}(1-\pi_{ij^{\prime}})}{\pi_{ij^{\prime}}(1-\pi_{ij})}\quad \text{subject to $\mathbf{x}_{ij}=\mathbf{x}_{ij^{\prime}}$ and $u_{ij}, u_{ij^{\prime}}\in [0,1]$ for all $i, j, j^{\prime}$.}
\end{equation*}
So we have under model (\ref{eqn: general interraction logit model}),
\begin{align*}
     \Gamma&=\max_{i, j, j^{\prime}} \max_{u_{ij}, u_{ij^{\prime}}\in [0,1]} \frac{\pi_{ij}(1-\pi_{ij^{\prime}})}{\pi_{ij^{\prime}}(1-\pi_{ij})}\\
     &=\max_{i, j, j^{\prime}}\max_{u_{ij}, u_{ij^{\prime}}\in [0,1]}\frac{\exp\{ g(\mathbf{x}_{ij})+\mathbf{\beta}^{T}\mathbf{x}_{ij}\times u_{ij}+ \gamma u_{ij}\}}{\exp\{ g(\mathbf{x}_{ij^{\prime}})+\mathbf{\beta}^{T}\mathbf{x}_{ij^{\prime}}\times u_{ij^{\prime}}+ \gamma u_{ij^{\prime}}\}}\\
      &=\max_{i, j, j^{\prime}}\max_{u_{ij}, u_{ij^{\prime}}\in [0,1]}\exp\{ (\mathbf{\beta}^{T}\mathbf{x}_{i}+\gamma)(u_{ij}-u_{ij^{\prime}}) \} \quad (\text{since $\mathbf{x}_{ij}=\mathbf{x}_{ij^{\prime}}=\mathbf{x}_{i}$})\\
      &=\max_{i, j, j^{\prime}}\exp( |\mathbf{\beta}^{T}\mathbf{x}_{i}+\gamma|) \\
      &=\max_{i=1,\dots,I}\exp(|\mathbf{\beta}^{T}\mathbf{x}_{i}+\gamma|).
\end{align*}
Therefore the desired equation holds.

\subsection*{Proof of Theorem~\ref{Theorem: Rosenbaum bounds are loose}}

\begin{proof}
Let $\mathbf{x}^{*}\in \{\mathbf{x}_{1}, \dots, \mathbf{x}_{I}\}$ be an observed covariate vector such that $\Gamma=\max_{i=1,\dots, I}\exp(|\mathbf{\beta}^{T}\mathbf{x}_{i}+\gamma|)=\exp(|\mathbf{\beta}^{T}\mathbf{x}^{*}+\gamma|)>1$. Since there exist two matched sets $i_{1}$ and $i_{2}$ such that $\mathbf{x}_{i_{1}}\neq \pm \mathbf{x}_{i_{2}}$, we have $\{\mathbf{x}_{1}, \dots, \mathbf{x}_{I}\}\setminus \{\mathbf{x}^{*}, -\mathbf{x}^{*}\}\neq \emptyset$. For any matched set $i$ such that $\mathbf{x}_{i}\neq \pm \mathbf{x}^{*}$, define the set
\begin{equation*}
    E_{i}=\Big \{ \mathbf{\beta}^{T} \in \mathbb{R}^{K}:\  \exp(|\mathbf{\beta}^{T}\mathbf{x}_{i}+\gamma|)=\exp(|\mathbf{\beta}^{T}\mathbf{x}^{*}+\gamma|) \text{\ and \ } \mathbf{\beta}^{T}\neq \mathbf{0} \Big \}.
\end{equation*}
Since $\mathbf{x}_{i}\neq \pm \mathbf{x}^{*}$, for any $\gamma \in \mathbb{R}$ we have
\begin{equation*}
    E_{i}=\Big \{ \mathbf{\beta}^{T} \in \mathbb{R}^{K}:\ \mathbf{\beta}^{T}(\mathbf{x}_{i}-\mathbf{x}^{*})=0 \text{\ and \ } \mathbf{\beta}^{T}\neq \mathbf{0}  \Big \} \cup \Big \{ \mathbf{\beta}^{T} \in \mathbb{R}^{K}:\ \mathbf{\beta}^{T}(\mathbf{x}_{i}+\mathbf{x}^{*})+2\gamma=0 \text{\ and \ } \mathbf{\beta}^{T}\neq \mathbf{0}  \Big \}
\end{equation*}
is a subset of $\mathbb{R}^{K}$ of Lebesgue measure zero. Let $E=\underset{i:\ \mathbf{x}_{i}\neq \pm \mathbf{x}^{*}}{\cup}E_{i}$, then $E$ is also a subset of $\mathbb{R}^{K}$ of Lebesgue measure zero. For any matched set $i$ such that $\mathbf{x}_{i}\neq \pm \mathbf{x}^{*}$, when $\mathbf{\beta}^{T}\neq \mathbf{0}$ and $\mathbf{\beta}^{T} \notin E$, note that
\begin{align*}
    \frac{\pi_{ij}(1-\pi_{ij^{\prime}})}{\pi_{ij^{\prime}}(1-\pi_{ij})}&=\frac{\exp\{ g(\mathbf{x}_{ij})+\mathbf{\beta}^{T}\mathbf{x}_{ij}\times u_{ij}+ \gamma u_{ij}\}}{\exp\{ g(\mathbf{x}_{ij^{\prime}})+\mathbf{\beta}^{T}\mathbf{x}_{ij^{\prime}}\times u_{ij^{\prime}}+ \gamma u_{ij^{\prime}}\}}\\
    &=\exp\{ (\mathbf{\beta}^{T}\mathbf{x}_{i}+\gamma)(u_{ij}-u_{ij^{\prime}}) \} \quad (\text{since $\mathbf{x}_{ij}=\mathbf{x}_{ij^{\prime}}=\mathbf{x}_{i}$})\\
    &\leq \exp(|\mathbf{\beta}^{T}\mathbf{x}_{i}+\gamma|) \quad (\text{since $u_{ij}, u_{ij^{\prime}}\in [0,1]$})\\
    &<\exp(|\mathbf{\beta}^{T}\mathbf{x}^{*}+\gamma|)\quad (\text{by definition of $E$})\\
    &=\Gamma,
\end{align*}
therefore setting $\Upsilon_{i}=\exp(|\mathbf{\beta}^{T}\mathbf{x}_{i}+\gamma|)$
we have, by symmetry,
\begin{equation*}
    \Gamma^{-1}<\exp(-|\mathbf{\beta}^{T}\mathbf{x}_{i}+\gamma|) \leq \frac{\pi_{ij}(1-\pi_{ij^{\prime}})}{\pi_{ij^{\prime}}(1-\pi_{ij})}\leq \exp(|\mathbf{\beta}^{T}\mathbf{x}_{i}+\gamma|)<\Gamma, \ \ \text{for all $j, j^{\prime}$.}
\end{equation*}
So the desired result follows.
\end{proof}

\subsection*{Proof of Theorem~\ref{thm: improved sensi model gamma}}

\begin{proof}
Under model (\ref{eqn: logit model with single interaction}), according to the definition of $\Gamma$ in the Rosenbaum bounds (\ref{eqn: sens model gamma}), we have
\begin{equation*}
    \Gamma=\max_{i, j, j^{\prime}}\frac{\pi_{ij}(1-\pi_{ij^{\prime}})}{\pi_{ij^{\prime}}(1-\pi_{ij})}\quad \text{subject to $\mathbf{x}_{ij}=\mathbf{x}_{ij^{\prime}}$ and $u_{ij}, u_{ij^{\prime}} \in [0,1]$ for all $i, j, j^{\prime}$},
\end{equation*}
where 
\begin{align*}
    \frac{\pi_{ij}(1-\pi_{ij^{\prime}})}{\pi_{ij^{\prime}}(1-\pi_{ij})}&=\frac{\exp\{ g(\mathbf{x}_{ij})+\widetilde{\beta}\ \widetilde{x}_{ij(k)} u_{ij}+ \gamma u_{ij}\}}{\exp\{ g(\mathbf{x}_{ij^{\prime}})+\widetilde{\beta}\ \widetilde{x}_{ij^{\prime}(k)} u_{ij^{\prime}}+ \gamma u_{ij^{\prime}}\}}\\
    &=\exp\{ (\widetilde{\beta}\ \widetilde{x}_{i(k)}+\gamma)(u_{ij}-u_{ij^{\prime}}) \} \quad (\text{since $\mathbf{x}_{ij}=\mathbf{x}_{ij^{\prime}}=\mathbf{x}_{i}$ and $\widetilde{x}_{ij(k)}=\widetilde{x}_{ij^{\prime}(k)}=\widetilde{x}_{i(k)}$})\\
    &\leq \exp(|\widetilde{\beta}\ \widetilde{x}_{i(k)}+\gamma|)\quad (\text{since $u_{ij}, u_{ij^{\prime}}\in [0,1]$}).
\end{align*}
Therefore, we have 
\begin{align*}
    \Gamma&=\max_{i, j, j^{\prime}}\frac{\pi_{ij}(1-\pi_{ij^{\prime}})}{\pi_{ij^{\prime}}(1-\pi_{ij})}\quad \text{subject to $\mathbf{x}_{ij}=\mathbf{x}_{ij^{\prime}}$ and $u_{ij}, u_{ij^{\prime}}\in [0,1]$ for all $i, j, j^{\prime}$}\nonumber \\ &=\max_{i=1,\dots,I}\exp(|\widetilde{\beta}\ \widetilde{x}_{i(k)}+\gamma|)\\
    &=\max \{ \exp(|\gamma|), \exp(|\widetilde{\beta}+\gamma|) \} \quad \text{\big(since $\widetilde{x}_{i(k)}=\frac{x_{i(k)}-\min_{i}x_{i(k)}}{\max_{i}x_{i(k)}-\min_{i}x_{i(k)}}\in [0,1]$\big)}.
\end{align*}
\begin{itemize}
    \item Case 1: $|\lambda|=|\frac{\widetilde{\beta}+\gamma}{\gamma}|\leq 1$. In this case we have $\Gamma=\max \{ \exp(|\gamma|), \exp(|\widetilde{\beta}+\gamma|) \}=\exp(|\gamma|)$. Therefore, we have
    \begin{align*}
    \frac{\pi_{ij}(1-\pi_{ij^{\prime}})}{\pi_{ij^{\prime}}(1-\pi_{ij})}&\leq \exp(|\widetilde{\beta}\ \widetilde{x}_{i(k)}+\gamma|)\\
    &=\exp\Big\{ |\gamma|\times \Big |\big(\frac{\widetilde{\beta}+\gamma}{\gamma}-1\big)\widetilde{x}_{i(k)}+1 \Big| \Big\}\\
    &=\Gamma^{|(\lambda-1) \widetilde{x}_{i(k)}+1|} \quad \Big(\text{since $\Gamma=\exp(|\gamma|)$ and $\lambda=\frac{\widetilde{\beta}+\gamma}{\gamma}$\Big).}
\end{align*} 
 Therefore, by symmetry we have
 \begin{equation*}
    \Gamma^{-|(\lambda-1) \widetilde{x}_{i(k)}+1|} \leq \frac{\pi_{ij}(1-\pi_{ij^{\prime}})}{\pi_{ij^{\prime}}(1-\pi_{ij})}\leq \Gamma^{|(\lambda-1) \widetilde{x}_{i(k)}+1|}, \ \ \text{for all $i, j, j^{\prime}$,}
 \end{equation*}
 and the above bounds are sharp in the sense that the upper bound can be achieved when $u_{ij}-u_{ij^{\prime}}=\text{sign}(\widetilde{\beta}\ \widetilde{x}_{i(k)}+\gamma)$, and the lower bound can be achieved when $u_{ij}-u_{ij^{\prime}}=-\text{sign}(\widetilde{\beta}\ \widetilde{x}_{i(k)}+\gamma)$, where we let $\text{sign}(x)$ equal $1$ if $x>0$, equal $-1$ if $x<0$, and equal $0$ if $x=0$. Since $ \widetilde{x}_{i(k)}\in [0,1]$ and $|\lambda|\leq 1$, we have $\Gamma^{|(\lambda-1) \widetilde{x}_{i(k)}+1|}\leq \Gamma$ and the equality holds if and only if at least one of the following three conditions holds: (a) $\lambda=1$; (b) $x_{i(k)}=\min_{i}x_{i(k)}$ (i.e., $\widetilde{x}_{i(k)}=0$); (c) $\lambda=-1$ and $x_{i(k)}=\max_{i}x_{i(k)}$ (i.e., $\widetilde{x}_{i(k)}=1$).

     \item Case 2: $|\lambda|=|\frac{\widetilde{\beta}+\gamma}{\gamma}|> 1$. In this case we have $\Gamma=\max \{ \exp(|\gamma|), \exp(|\widetilde{\beta}+\gamma|) \}=\exp(|\widetilde{\beta}+\gamma|)$. So we have
    \begin{align*}
    \frac{\pi_{ij}(1-\pi_{ij^{\prime}})}{\pi_{ij^{\prime}}(1-\pi_{ij})}&\leq \exp(|\widetilde{\beta}\ \widetilde{x}_{i(k)}+\gamma|)\\
    &=\exp\Big\{ |\widetilde{\beta}+\gamma|\times \Big |\big(1-\frac{\gamma}{\widetilde{\beta}+ \gamma}\big)\widetilde{x}_{i(k)}+\frac{\gamma}{\widetilde{\beta}+ \gamma} \Big| \Big\}\\
    &=\Gamma^{|(1-\lambda^{-1}) \widetilde{x}_{i(k)}+\lambda^{-1}|} \quad \Big(\text{since $\Gamma=\exp(|\widetilde{\beta}+\gamma|)$ and $\lambda=\frac{\widetilde{\beta}+\gamma}{\gamma}$\Big).}
\end{align*} 
 Therefore, by symmetry we have
 \begin{equation*}
    \Gamma^{-|(1-\lambda^{-1}) \widetilde{x}_{i(k)}+\lambda^{-1}|} \leq \frac{\pi_{ij}(1-\pi_{ij^{\prime}})}{\pi_{ij^{\prime}}(1-\pi_{ij})}\leq \Gamma^{|(1-\lambda^{-1}) \widetilde{x}_{i(k)}+\lambda^{-1}|}, \ \ \text{for all $i, j, j^{\prime}$,}
    \end{equation*}
\end{itemize}
and the above bounds are sharp, which is similar to the argument in Case 1. Since $ \widetilde{x}_{i(k)}\in [0,1]$ and $|\lambda|> 1$, we have $\Gamma^{|(1-\lambda^{-1}) \widetilde{x}_{i(k)}+\lambda^{-1}|}\leq \Gamma$ and the equality holds if and only if $x_{i(k)}=\max_{i}x_{i(k)}$ (i.e., $\widetilde{x}_{i(k)}=1$).

The desired result follows from combining the arguments in Case 1 and Case 2.
\end{proof}

\subsection*{Proof of Corollary~\ref{corollary: dummy improved sensi model gamma}}

\begin{proof}
Consider the $\Gamma_{\lambda, i}$ defined in Theorem~\ref{thm: improved sensi model gamma}.
\begin{itemize}
    \item Case 1: $|\lambda|=1$. In this case, we have $\Gamma_{\lambda, i}=\Gamma^{|(\lambda-1) \widetilde{x}_{i(k)}+1|}=\Gamma$ for $x_{i(k)}\in \{0,1\}$.
    \item Case 2: $|\lambda|<1$. In this case, if $x_{i(k)}=0$, we have $\Gamma_{\lambda, i}=\Gamma^{|(\lambda-1) \widetilde{x}_{i(k)}+1|}=\Gamma$. If $x_{i(k)}=1$, we have $\Gamma_{\lambda, i}=\Gamma^{|(\lambda-1) \widetilde{x}_{i(k)}+1|}=\Gamma^{|\lambda|}$.
    \item Case 3: $|\lambda|>1$. In this case, if $x_{i(k)}=1$, we have $\Gamma_{\lambda, i}=\Gamma^{|(1-\lambda^{-1}) \widetilde{x}_{i(k)}+\lambda^{-1}|}=\Gamma$. If $x_{i(k)}=0$, we have $\Gamma_{\lambda, i}=\Gamma^{|(1-\lambda^{-1}) \widetilde{x}_{i(k)}+\lambda^{-1}|}=\Gamma^{1/|\lambda|}$.
\end{itemize}
Then the desired result follows immediately from applying Theorem~\ref{thm: improved sensi model gamma}.
\end{proof}

\subsection*{Proof of Corollary~\ref{corollary: smaller worst-case p-value}}

\begin{proof}
The proof follows from a direct adjustment of the proof of Proposition 13 in \citet{rosenbaum2002observational}. For each fixed $\widetilde{\beta}, \gamma, u_{ij}$, $i=1,\dots, I$ and $j=1, 2$, the test statistic $T_{\text{ss}}$ is the sum of $I$ independent random variables, where the $i$th variable equals $d_{i}$ with probability 
\begin{equation*}
    p_{i}=\frac{c_{i1}\exp\{(\widetilde{\beta}\ \widetilde{x}_{i(k)}+\gamma)(u_{i1}-u_{i2})\}+c_{i2}}{1+\exp\{(\widetilde{\beta}\ \widetilde{x}_{i(k)}+\gamma)(u_{i1}-u_{i2})\}},
\end{equation*}
and equals $0$ with probability $1-p_{i}$. Note that from the proof of Theorem 2, we have $\Gamma_{\lambda, i}=\exp\{|\widetilde{\beta}\ \widetilde{x}_{i(k)}+\gamma|\}$. Following the proof of Proposition 13 in \citet{rosenbaum2002observational}, the upper bound distribution $ \text{pr}(\sum_{i=1}^{I}\widetilde{T}_{\Gamma, \lambda, i} \geq t \mid  \mathcal{F}, \mathcal{Z})$ is the distribution of $T_{\text{ss}}$ when $u_{ij}=c_{ij}$ if $\widetilde{\beta}\ \widetilde{x}_{i(k)}+\gamma\geq 0$ and $u_{ij}=1-c_{ij}$ if $\widetilde{\beta}\ \widetilde{x}_{i(k)}+\gamma< 0$, resulting in the desired 
\begin{align*}
    \widetilde{p}_{\lambda,i}&=\left\{
\begin{array}{ll}
      0 & \quad \text{if\ $c_{i1}=c_{i2}=0$}, \\
      1 & \quad  \text{if\ $c_{i1}=c_{i2}=1$}, \\
      \frac{\exp\{|\widetilde{\beta}\ \widetilde{x}_{i(k)}+\gamma|\}}{1+\exp\{|\widetilde{\beta}\ \widetilde{x}_{i(k)}+\gamma|\}} & \quad \text{if \ $c_{i1}\neq c_{i2}$}.
\end{array}
\right. \\
&=\left\{
\begin{array}{ll}
      0 & \quad \text{if\ $c_{i1}=c_{i2}=0$}, \\
      1 & \quad  \text{if\ $c_{i1}=c_{i2}=1$}, \\
      \frac{\Gamma_{\lambda, i}}{1+\Gamma_{\lambda, i}} & \quad \text{if \ $c_{i1}\neq c_{i2}$}.
\end{array}
\right. 
\end{align*}
Applying Theorem~\ref{thm: improved sensi model gamma}, the inequality $\text{pr}(\sum_{i=1}^{I}\widetilde{T}_{\Gamma, \lambda, i} \geq t \mid  \mathcal{F}, \mathcal{Z})\leq  \text{pr}(\sum_{i=1}^{I}\overline{T}_{\Gamma, i} \geq t \mid  \mathcal{F}, \mathcal{Z})$ holds for all $t$, $\Gamma> 1$ and $\lambda \in \mathbb{R}$.
\end{proof}

\bibliographystyle{apalike}
\bibliography{paper-ref}

\end{document}